\documentclass[a4paper,twoside]{article}

\usepackage{epsfig}
\usepackage{subcaption}
\usepackage{calc}
\usepackage{amssymb}
\usepackage{amstext}
\usepackage{amsmath}
\usepackage{amsthm}
\usepackage{multicol}
\usepackage{pslatex}
\usepackage{apalike}
\usepackage{algorithm2e}
\usepackage[bottom]{footmisc}
 \usepackage{booktabs}
\usepackage{graphicx}
\usepackage{tabularx}
\usepackage{makecell}
\usepackage{rotating}
\usepackage{multirow}
\usepackage{color, colortbl}
\usepackage{smartdiagram}
\usepackage{pgfplots}
\pgfplotsset{compat=1.18}
\usepackage{balance}

\usesmartdiagramlibrary{additions}
\definecolor{Gray}{gray}{0.8}
\usepackage{tikz}
\usetikzlibrary{arrows.meta, positioning}

\usepackage{SCITEPRESS}     % Please add other packages that you may need BEFORE the SCITEPRESS.sty package.

\begin{document}

\title{Modelling GDPR-based Privacy Requirements with Software Engineering Diagrams: A Systematic Literature Review}

\author{\authorname{Evangelia Vanezi, Georgia Kapitsaki and Anna Philippou}
\affiliation{Department of Computer Science, University of Cyprus}
\email{\{vanezi.evangelia, gkapi, annap\}@ucy.ac.cy}
}


\abstract{The application of the General Data Protection Regulation (GDPR) has significantly affected privacy requirements elicitation, modelling, and verification in Software Engineering (SE). 
One of the affected areas is requirements visualisation through modelling diagrams, which plays a crucial role in ensuring privacy compliance, as functional system requirements should be integrated with GDPR-based privacy requirements. 
We present a systematic literature review on how SE diagrams have been employed to capture and integrate GDPR-based privacy requirements into software system design. The study aims to identify the existing research landscape, existing gaps, and directions for future work.
Following a rigorous search protocol and addressing two research questions, 18 primary studies published between 2017 and 2025 were selected, analysed, and categorised based on (i) the diagram types used, and (ii) the GDPR principles or rights addressed.
The findings highlight the need for inter-diagram integration, full lifecycle traceability mechanisms, tool support, and automated compliance checking.  
}

\onecolumn \maketitle \normalsize \setcounter{footnote}{0} \vfill

\section{\uppercase{Introduction}}
The European Union's General Data Protection Regulation (GDPR)~\cite{eugdpr} was voted in 2016 and has been applied since May 25$^{th}$, 2018, enforcing major changes in the way organisations operate in terms of personal data collection, processing, and management.
For software systems, this change directly impacts the way privacy requirements are elicited, specified, and verified throughout the development lifecycle.
To provide the GDPR-imposed data subject rights and comply with GDPR provisions and liabilities for data processors and data controllers, software systems design should incorporate GDPR-based privacy requirements alongside functional requirements. 

Since the GDPR was introduced, numerous studies have explored ways to translate regulation obligations into actionable software requirements, aiming to incorporate GDPR compliance into software system design and development.
%
%The GDPR imposes rights for data owners and liabilities for data processors and data controllers, bearing strict penalties for non-compliance. 
%The regulation was voted on in 2016 and, since then, many attempts have been made to incorporate GDPR compliance into software system design and development. 
Works have studied the interpretation of GDPR articles into system functionality and their incorporation into software design~\cite{vanezi2019gdpr}, the GDPR compliance and completeness of privacy policies~\cite{torre2020ai,vanezi2021complicy,contissa2018claudette}, while others attempted to provide software users with a better experience around their GDPR-imposed privacy rights~\cite{vanezi2023saving}, study and enhance users' perception on GDPR-based privacy policies~\cite{theophilou2025ohkey,vanezi2025making}, or connect visual GDPR representations with formal languages~\cite{vanezi2020dialogop}.
Nevertheless, ensuring traceable, design-level compliance remains a key challenge for both researchers and practitioners in Software Engineering (SE).

Requirements visualisation through software modelling diagrams, such as Unified Modelling Language (UML) diagrams~\cite{booch1996unified} or Data Flow Diagrams (DFDs)~\cite{tao1991formal}, can play a crucial role in bridging the gap between abstract regulation principles and concrete system design, towards ensuring privacy compliance. 
When properly extended or annotated, such diagrams can help software designers model how personal data are collected, used, or deleted, making GDPR compliance explicit and verifiable at design time.

While several works have studied privacy and the GDPR from broader requirements engineering perspectives~\cite{morel2020sok,negri2024understanding,saltarella2021privacy}, no prior review has systematically examined how widely used modelling diagrams are employed to capture and integrate GDPR-based privacy requirements. Existing surveys largely overlook this visual modelling dimension, focusing instead on textual or process-based approaches.

To address this gap, we conducted a Systematic Literature Review (SLR) to investigate how SE diagrams %Unified Modelling Language and Data Flow Diagrams 
have been employed to model GDPR-based privacy requirements in software system design. The study identifies which diagram types are used (Data Flow Diagrams, UML Class, Sequence, Use Case, Activity, and Collaboration Diagrams), and which GDPR notions are addressed through them (Privacy-by-Design, Purpose Limitation, Data Minimisation, Explicit and Informed Consent, Accountability, Storage Limitation, and the Rights to Erasure and Access), recognising gaps and challenges. 

Our SLR follows a rigorous
search protocol aiming to address two research questions. As a result, 18 relevant works published between 2017 and 2025 were selected, analysed, and categorised. 
Our synthesized findings reveal the need for 
inter-diagram integration, full lifecycle traceability mechanisms, tool support, and automated compliance checking.
%more comprehensive frameworks connecting privacy requirements with formal representations towards automated compliance verification and requirements traceability.

The remainder of this paper is organized as follows. Section~\ref{related_work} presents related work, Section~\ref{SLR_section} discusses the SLR Protocol, and Section~\ref{results} presents the results. 
Section~\ref{discussion_section} discusses the findings, gaps, opportunities, and threats to validity, while Section~\ref{conclusions_section} concludes the paper.

\section{Related work}
\label{related_work}
There have been several review studies in the area of privacy requirements collection and modelling. 
Morel et al. systematically studied
different means of expressing privacy policies as the main way to obtain information related to the collection and processing of personal data~\cite{morel2020sok}. Riva et al. examined existing methodologies for the design of privacy-aware systems~\cite{riva2020sok}, investigating the extent to which these methodologies align with the GDPR and Privacy-by-Design principles, address different levels of system design concerns, and demonstrate their suitability for the intended purpose.

In~\cite{alpers2018identifying}, approaches for modelling privacy from business and software engineering perspectives are examined. The authors' key finding is that, at the time, no comprehensive modelling approach fully addressed the needed aspects and perspectives. Negri-Ribalta et al.~\cite{negri2024understanding} aimed to identify research in requirements engineering that supports compliance with regulatory data protection requirements. Their analysis focused on key trends, such as year of publication, publication venue, type of research, interdisciplinarity in the authors' background, GDPR focus, and type of proposal. 

Furthermore, Saltarella et al. presented a systematic literature review identifying useful guidelines to support the development of GDPR-compliant software~\cite{saltarella2021privacy}, while Aberkane et al. provide a
systematic mapping study on the intersection of GDPR, natural language processing, and requirements engineering~\cite{aberkane2021exploring}.

While several reviews have addressed GDPR and privacy in requirements engineering, there is no work focusing specifically on the diagrammatic visualisations of GDPR-based privacy requirements, nor on how these methods bridge regulatory compliance and system-level design, as addressed in the current work.

\section{SLR Protocol}
\label{SLR_section}
To systematically examine how software modelling diagrams are employed for capturing and integrating GDPR-based privacy requirements, we conducted a Systematic Literature Review following a rigorous protocol drafted based on the guidelines by Kitchenham et al.~\cite{keele2007guidelines} for Systematic Literature Reviews in software engineering research. As a first step, we defined our two Research Questions (RQs). Based on them, we developed our search terms and phrases, which were used to query a selected set of digital libraries. 
Subsequently, the retrieved studies were filtered using predefined inclusion criteria, duplicates were removed, and additional relevant works were incorporated through snowballing. 
A quality assessment checklist was then applied for the final selection. 

The process resulted in 18 primary studies, which were analysed in depth to extract important data and synthesize the results to address the two RQs. The following subsections describe the review protocol in detail.

\subsection{Research Questions} 
\label{rqs}
We defined two Research Questions to guide this SLR, aiming to identify both the diagram types employed and the GDPR-related privacy notions addressed in the relevant literature, as follows: 
\begin{itemize}
    \item \textbf{RQ1}. What types of SE diagrams have been used in the literature to integrate GDPR-based privacy requirements into software system design?
    \item \textbf{RQ2}. Which GDPR privacy notions have been incorporated into software system requirements and design with the use of SE diagrams? 
\end{itemize}
Following the specification of the RQs, we developed our review protocol as described in the following sub-sections.
\subsection{Search Phrases} 
We developed a set of search phrases (SPs) following the guidelines provided by Kitchenham and Charters~\cite{keele2007guidelines}:
(1) extracting the key concepts from each RQ and identifying relevant keywords; (2) including synonyms identified from the literature; (3) using Boolean operators (AND, OR) to combine keywords and synonyms into search phrases. The final set of phrases is presented in Table~\ref{tab:search-phrases}.
\begin{table}[ht]
\caption{Search phrases used in the SLR.}
\centering
\resizebox{0.5\textwidth}{!}{%
\begin{tabular}{|l|l|} \hline
\textbf{SP1} &  ``UML diagrams" AND GDPR \\
\hline
\textbf{SP2} &  ``Sequence diagrams" AND GDPR \\
\hline
\textbf{SP3} & ``Data flow diagrams" AND GDPR \\
\hline
\textbf{SP4} & ``Use case diagrams" AND GDPR \\
\hline
\textbf{SP5} & ``Class diagrams" AND GDPR \\
\hline
\textbf{SP6} & UML AND GDPR \\
\hline
\textbf{SP7} & ``Software Engineering diagrams" AND GDPR \\
\hline
\textbf{SP8} & ``Activity diagrams" AND GDPR \\
\hline
\textbf{SP9} & ``Context diagrams" AND GDPR \\
\hline
\textbf{SP10} & ``Component diagrams" AND GDPR \\
\hline
\textbf{SP11} & ``Object diagrams" AND GDPR \\
\hline
\textbf{SP12} & ``Deployment diagrams" AND GDPR \\
\hline
\textbf{SP13} & ``State diagrams" AND GDPR \\
\hline
\textbf{SP14} & ``Requirement diagrams" AND GDPR \\
\hline
\textbf{SP15} & ``Analysis diagrams" AND GDPR \\
\hline
\textbf{SP16} & ``Design diagrams" AND GDPR \\
\hline
\textbf{SP17} & ``Goal models" AND GDPR \\
\hline
\textbf{SP18} & BPMN AND GDPR \\
\hline
\end{tabular}
}
\label{tab:search-phrases}
\end{table}

The key concepts include some of the most commonly used SE diagrams, as well as more generic terms such as ``Analysis Diagrams", or ``Software Engineering Diagrams", in order to capture works that employ diagrams not explicitly listed in the search phrases. 

% \begin{itemize}
%     \item \textbf{SP1}. ``UML diagrams" AND GDPR
%     \item \textbf{SP2}. ``Sequence diagrams" AND GDPR
%     \item \textbf{SP3}. ``Data flow diagrams" AND GDPR
%     \item \textbf{SP4}. ``Use case diagrams" AND GDPR
%     \item \textbf{SP5}. ``Class diagrams" AND GDPR
%     \item \textbf{SP6}. UML AND GDPR
%     \item SP7. ``Software Engineering Diagrams" AND GDPR
%     \item SP8. Activity Diagrams
%     \item SP9. Context Diagrams
%     \item SP10. Component Diagrams
%     \item SP11. Object diagram
%     \item SP12. Deployment diagrams
%     \item SP13. State diagram
% \end{itemize}
%%
\subsection{Digital Libraries and Data Sources} 
Aiming to perform a comprehensive and
structured search, we conducted our search in the following widely used scientific and general-purpose databases: \textbf{Science Direct}, \textbf{ACM Digital Library}, \textbf{IEEE Xplore}, and \textbf{Google Scholar}. 
These sources were selected to ensure broad coverage of research related to our areas of interest.

Table~\ref{tab:search-results} presents the number of publications returned for each search phrase in each digital library.

\begin{table}[ht]
\caption{Number of publications per source (parentheses indicate publications passing the inclusion criteria).}
\centering
\resizebox{0.46\textwidth}{!}{%
\begin{tabular}{|l|c|c|c|c|c|c|}
\hline
  \textbf{Source} & \multicolumn{6}{c|}{\textbf{Search Phrases}} \\
\hline
\rowcolor{Gray}  & \textbf{SP1} & \textbf{SP2} & \textbf{SP3} & \textbf{SP4} & \textbf{SP5} & \textbf{SP6} \\
  \hline 
 Science Direct & 19 (2) & 72 (1) & 45 (2) & 17 (0) & 31 (0) & 127 (4) \\  \hline 
  ACM & 50 (0) & 47 (0) & 36 (3) & 16 (0) & 58 (1) & 216 (7)\\   \hline 
IEEE Xplore & 1 (1) & 1 (0) & 1 (0)& 0 (0) & 1 (0) & 7 (2)\\  \hline 
Google Scholar & 842 (12) & 395 (6) & 801 (18) & 241 (3) & 336 (9) & 3220 (12) \\ 
\hline 
\rowcolor{Gray}  & \textbf{SP7} & \textbf{SP8} & \textbf{SP9} & \textbf{SP10} & \textbf{SP11} & \textbf{SP12} \\
  \hline 
 Science Direct & 0 (0) & 25 (0) & 9 (0) & 13 (0) & 3 (0) & 4 (0) \\  \hline 
  ACM & 0 (0) & 28 (1) & 1 (0) & 16 (0) & 8 (0) & 6 (0)\\  \hline 
IEEE Xplore & 22 (2) & 0 (0) & 0 (0)& 0 (0) &  5 (1) & 0 (0)\\  \hline 
Google Scholar & 6 (0) & 360 (11) & 49 (3) &  91 (6) & 29 (1) & 133 (2) \\ 
\hline
\rowcolor{Gray}  & \textbf{SP13} & \textbf{SP14} & \textbf{SP15} & \textbf{SP16} & \textbf{SP17} & \textbf{SP18}  \\
  \hline 
Science Direct & 8 (1) & 1 (0) & 12 (0) & 6 (1) & 17 (2)  & 63 (1) \\  \hline 
  ACM & 39 (1) & 0 (0) & 3 (0) & 17 (0) & 2 (0) & 98 (1) \\   \hline 
IEEE Xplore & 0 (0) & 0 (0) & 0 (0)& 0 (0) & 1 (0)  & 5 (1) \\  \hline 
Google Scholar & 81 (3) & 7 (0) & 22 (0) & 57 (1) & 173 (3) & 2020 (10)  \\
\hline
\end{tabular}%
}
\label{tab:search-results}
\end{table}

\subsection{Screening and Selection of Primary Studies}
After collecting the initial search results, we first removed duplicate publications returned by different search phrases or data sources.
We then formulated a list of inclusion criteria (IC1-IC6), through which we filtered the retrieved articles to determine the primary studies to be included in the review. Publications that did not meet all criteria were excluded.

\textbf{Inclusion Criteria (ICs):}
\begin{itemize}
    \item \textbf{IC1.} Language of the publication: English.
    \item \textbf{IC2.} Acceptance of the publication: Peer-reviewing process. 
    \item \textbf{IC3.} Type of the publication: Scientific publications (journal articles, conference papers, academic thesis/dissertations). 
    \item \textbf{IC4.} Year of publication: 2016 to 2025, covering the period since GDPR was announced. 
    \item \textbf{IC5.} Relevance of the publication to our RQs: Primary studies describing the use of SE diagrams %UML diagrams or DFDs 
    to support software system design incorporating GDPR-based privacy requirements.
    \item \textbf{IC6.} Accessibility of the publication: The full text should be available.
\end{itemize}
We filtered the articles by screening their titles, abstracts, and, when necessary, parts of the full text. 
All articles were independently screened and reviewed by two of the authors, both computer scientists with experience in GDPR. In cases of initial disagreement, all three authors discussed the inclusion decision and reached consensus through joint re-evaluation, with the intervention of the third author who had not previously been involved in reviewing the work and thus remained impartial. %a third senior researcher intervened and took the final decision. 

%\paragraph{Exclusion.}
Most ICs were straightforward to address. Regarding relevance (criterion IC5), some works were observed to use SE diagrams only as auxiliary support to their methodology, rather than as the primary means for modelling systems with GDPR-based privacy requirements. Our review focuses specifically on studies that model software systems through SE diagrams while explicitly incorporating GDPR-based privacy requirements at the requirements or design level. 
Other relevant works were considered out of scope, such as studies that use goal-oriented models to represent privacy regulations rather than modelling the systems themselves~\cite{rabinia2018fol}, works focusing on concerns such as conflicts between security, data minimisation, and fairness~\cite{ramadan2020semi}, threat modelling and impact assessment~\cite{sion2019architectural}, % GDPR compliance checking~\cite{capodieci2025enhancing}, 
or approaches aimed at detecting data leakages~\cite{pullonen2017pe,belluccini2020verification}. Such works, while valuable, do not align with the specific focus of this SLR and were therefore excluded specifically following evaluation with inclusion criterion IC5.
%For example, some works were observed to use SE diagrams as auxiliary support of their methodology rather than as the basic means for privacy-aware design. After thorough joint re-evaluation, these works were excluded. 

Table~\ref{tab:search-results} shows the number of works that passed the inclusion criteria from the results retrieved for each search phrase-library combination (in parenthesis). 

%Subsequently, we added a few more publications by searching the references and citations of the articles collected so far, to identify other relevant material (snowballing). 
We complemented the databases search with snowballing from the references of the selected studies. Studies collected, were also filtered against the inclusion criteria.
Then, a quality assessment checklist was drafted, through which all remaining papers were screened:
\begin{enumerate}
    \item \textbf{``Are the study objectives clearly stated?"}
    \item \textbf{``Does the work provide a clear contribution to the field?"}
    \item \textbf{``Is the methodology adequately described?"}
\end{enumerate}

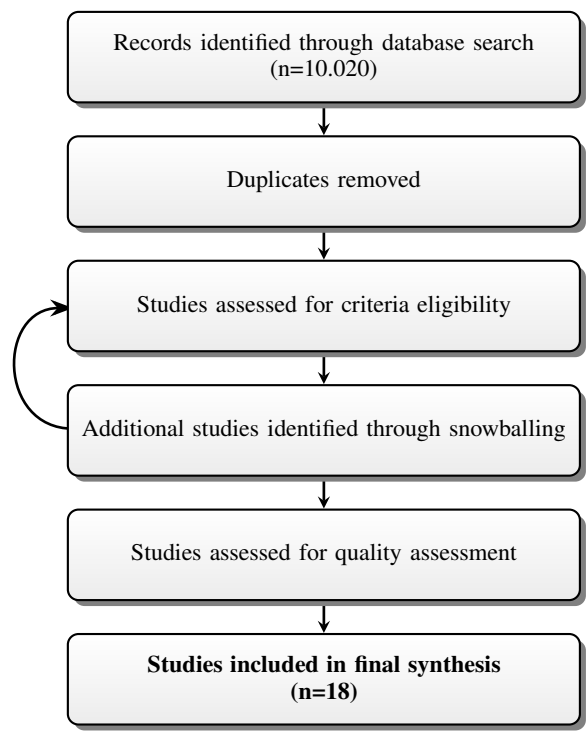
\begin{figure}[h!]
\centering
\begin{tikzpicture}[remember picture]
\node (diagram) {
\smartdiagram[flow diagram:vertical]{
  Records identified through database search\\(n=10.020),
  Duplicates removed,
  Studies assessed for criteria eligibility,
  Additional studies identified through snowballing,
 Studies assessed for quality assessment,
  \textbf{Studies included in final synthesis\\(n=18)}
}
};
\end{tikzpicture}
\caption{PRISMA-style flow diagram illustrating the study selection process.}
\label{fig:prisma-flow}
\end{figure}

Applying these procedures resulted in 18 primary studies included in the final synthesis.
Figure~\ref{fig:prisma-flow} illustrates the overall study selection process.

\smartdiagramset{
  uniform color list = {gray!15} for 6 items, % 5 blue nodes
  back arrow disabled = true,
  module minimum width = 3.5cm,
  module minimum height = 1.2cm,
  text width = 6.5cm,
  font = \small,
  border color = black,
  arrow line width = 1pt,
  uniform arrow color = true,
  arrow color = black
}
\begin{tikzpicture}[remember picture, overlay]
\draw[-{Stealth[length=3mm]}, line width=1pt]
  ([xshift=-3.3cm,yshift=-0.8cm]diagram.center)
  .. controls +(-1.0,0) and +(-1.0,0) ..
  ([xshift=-3.3cm,yshift=0.8cm]diagram.center);
\end{tikzpicture}

\subsection{Data Extraction and Synthesis}
To answer the research questions and reveal the overall findings, we analysed the collected articles by extracting the following data items from each study:
\begin{enumerate}
    \item \textbf{Bibliographic data (authors, affiliation, year, and publication venue);}
    \item \textbf{Objectives of the study (e.g., requirements elicitation, design, verification, etc.);}
    \item \textbf{A brief summary of the study;}
    \item \textbf{Diagram type(s) used;}
    \item \textbf{GDPR notion(s) addressed;}
    \item \textbf{Methodological approach (extension or case study).}
\end{enumerate}
The data extraction process was straightforward and did not require additional interpretation or inference. %Guidelines by Kitchenham and Charters~\cite{keele2007guidelines} were followed. 
The data were extracted by the first author using a standardized spreadsheet template. To reduce bias, the second author independently cross-checked the extracted data. 

Following data extraction, the data items were reviewed by all three authors to answer the research questions and to identify emerging trends and limitations. Data were synthesized through thematic grouping to identify patterns across studies, as well as descriptive statistics. Qualitative insights informed the categorisation, while quantitative data supported the construction of the tables presented in Section~\ref{results}. 

\section{Results}
\label{results}

The synthesized results are presented in alignment with the research questions introduced in Section~\ref{rqs}. We first provide an overview of the selected studies and their classification into methodological categories (Section~\ref{overview}), followed by detailed analyses addressing RQ1 and RQ2 (Sections~\ref{rq1} and ~\ref{rq2}).

\subsection{Overview and Study Categories}
\label{overview}
Our final set of primary studies includes 18 works published between 2017 and 2025, with a near-even split between the two periods (10 studies in 2017–2020 and 8 in 2021–2025), indicating sustained research activity throughout the years (Table~\ref{tab:SLR_periods}). Notably, two studies predate GDPR enforcement (2018), reflecting anticipatory work during the transition period following its announcement.
\begin{table}[ht]
\caption{Number of works per period.}
\centering
%\resizebox{0.47\textwidth}{!}{%
\begin{tabular}{lcc}
\toprule
Period & 2017--2020 & 2021--2025 \\
\midrule
Number of studies & 10 & 8 \\
\bottomrule
\end{tabular}
%}
\label{tab:SLR_periods}
\end{table}

Based on data item 6 (methodological approach), the studies were classified into two main categories:
\begin{itemize}
    \item \textbf{Category 1: Works that design privacy-compliant functionality for a specific system using standard SE diagrams. %UML diagrams or DFDs.
    } These works focus on integrating privacy principles into system design for a specific application domain (e.g., e-learning, IoT, healthcare) or even a single system. They use traditional SE diagrams %UML diagrams and DFDs 
    without modifying their structure, while applying privacy best practices (e.g., data minimisation) within the system functionality.
    \item \textbf{Category 2: Works that propose an extension of SE diagrams %UML diagrams or DFDs 
    to model any system with privacy compliance.} These works modify SE diagrams %UML or DFDs 
    by adding new privacy-focused elements (e.g., privacy annotations or constraints). They aim to create generalised modelling techniques applicable to a wide range of systems.
\end{itemize}
The distribution of studies across categories is shown in Table~\ref{tab:SLR_categories}. 
While Category 1 approaches demonstrate how standard diagrams can be applied to model GDPR-compliant functionality in specific application contexts, their applicability is limited to the systems under study. In contrast, Category 2 approaches aim to extend or generalise modelling notations, enabling the representation of GDPR-based privacy requirements across a broader range of systems and domains, at the cost of increased modelling complexity and a potentially steeper learning curve for engineers.

\begin{table}[ht]
\caption{Works per methodological category.}
\centering
\resizebox{0.47\textwidth}{!}{%
\begin{tabular}{l|c|l}
\hline
\rowcolor{Gray}  \textbf{Cat.} & \textbf{\#} & \textbf{Papers} \\
\hline
 1 & 4 &
 \makecell[l]{
\cite{mougiakou2017based}, 
\cite{kammuller2019designing},\\ \cite{cambronero2024towards},  
 \cite{vanezi2019gdpr}} \\
  \hline 
  2 & 14 &
 \makecell[l]{
\cite{pedroza2021model}, 
\cite{rahman2017petri}, 
\cite{veseli2019engineering},\\
\cite{ahmadian2018extending}, 
\cite{peyrone2022formal2},\\
%\cite{sion2018interaction}, 
%\cite{wuyts2020linddun},
\cite{ye2023mbipv}, 
%\cite{torre2020model}, 
%\cite{torre2021modeling}, 
%\cite{mai2018modeling},
\cite{alshareef2022precise},\\
\cite{antignac2018privacy}, 
\cite{ahmadian2019privacy}, \\
\cite{angergaard2022realizing},
\cite{alshareef2021refining},\\  
\cite{alshareef2021transforming},\\
\cite{ferreyra2020pdp}, \cite{vanezi2025privacy}
} 
\\
\hline 
\end{tabular}%
}
\label{tab:SLR_categories}
\end{table}

\begin{figure}[t]
\centering
\begin{tikzpicture}
\begin{axis}[
    ybar stacked,
    bar width=7pt,
    width=\columnwidth,
    height=5.2cm,
    ymin=0,
    ymax=4,
    ylabel={Number of publications},
    xlabel={Year},
    symbolic x coords={2017,2018,2019,2020,2021,2022,2023,2024,2025},
    xtick=data,
    x tick label style={font=\small},
    ytick={0,1,2,3,4},
    legend style={
        at={(0.5,1.02)},
        anchor=south,
        legend columns=2,
        draw=none,
        font=\small
    },
    nodes near coords,
    every node near coord/.append style={font=\scriptsize},
    enlarge x limits=0.06
]

\addplot coordinates {
    (2017,1) (2018,0) (2019,2) (2020,0) (2021,0)
    (2022,0) (2023,0) (2024,1) (2025,0)
};

\addplot coordinates {
    (2017,1) (2018,2) (2019,2) (2020,1) (2021,3)
    (2022,3) (2023,1) (2024,0) (2025,1)
};

\legend{Category 1, Category 2}
\end{axis}
\end{tikzpicture}
\caption{Temporal distribution of the selected studies by methodological category. Category 1 includes works that apply standard SE diagrams to specific systems, whereas Category 2 includes works that extend SE diagrams for privacy-compliant modelling.}
\label{fig:temporal}
\end{figure}
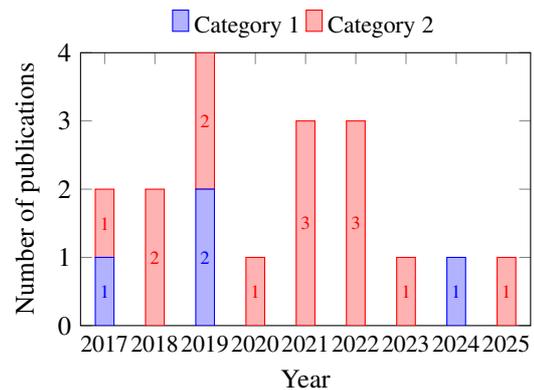

Figure~\ref{fig:temporal} shows the temporal distribution of the selected studies by methodological category. The plot indicates sustained research activity across the review period, with Category 2 studies dominating most years, while Category 1 studies appear more sporadically.

%%
% \subsubsection{Patterns}
% Five different patterns have emerged from the reviewed papers:
%%

\subsection{RQ1 – Diagram Types Used for GDPR-based Privacy Modelling}
\label{rq1}
%\noindent\rule{\linewidth}{0.4pt}
\fbox{\begin{minipage}{0.45\textwidth}
\textit{\textbf{RQ1.} What types of %UML diagrams or DFDs 
SE diagrams have been used in the literature to integrate GDPR-based privacy requirements into software system design?}
\end{minipage}}
\\
\\
%\noindent\rule{\linewidth}{0.4pt}
Data Flow Diagrams (DFDs) and different UML-based diagrams have been employed in the studies explored. Specifically, across both categories, six diagram families were identified: DFDs / Privacy-Aware DFDs (PA-DFDs), UML Class, Sequence, Use Case, Activity, and Collaboration diagrams. 
Category 1 works apply Use Case, Collaboration, and Sequence Diagrams in concrete case studies, while Category 2 works focus primarily on DFDs or UML extensions (e.g., privacy-aware profiles, PA-DFDs), with the exception of Collaboration Diagrams.
Below, we discuss the different diagram types and the respective works.

\paragraph{$\bullet$ \textbf{Data Flow Diagrams (DFDs) / Privacy-Aware DFDs (PA-DFDs).}} 
A number of works study Privacy-Aware Data Flow Diagrams (PA-DFDs) either implementing, extending, or refining them, often building on each other. 
The works of~\cite{antignac2018privacy} and~\cite{alshareef2021transforming} introduce approaches for automated transformation of DFDs into PA-DFDs. 
Specifically, ~\cite{antignac2018privacy} introduces PA-DFDs by defining privacy-specific elements in DFDs, and proposes an approach to recognize privacy-sensitive hotspots in DFDs for applying automated transformation, by using predefined transformation rules. Moreover,~\cite{alshareef2021transforming} proposes a semi-automated algorithm to systematically convert DFDs to PA-DFDs enforcing privacy constraints, and an implementation of their transformation in Python.

A refinement is proposed in~\cite{alshareef2021refining}, which ensures that privacy properties remain valid across system changes, focusing on design evolution.
The works of~\cite{rahman2017petri} and~\cite{alshareef2022precise} introduce theoretical foundations for PA-DFDs towards formal verification. In particular, in~\cite{rahman2017petri} Petri nets are employed to provide formal semantics for PA-DFDs, aiming to verify privacy properties compliance. The work presents an algorithm to convert PA-DFDs into Petri nets.
In~\cite{alshareef2022precise}, PA-DFDs are extended with purpose labels, added on data flows to label them according to the intended data usage. Automated checking of unauthorized data usage is proposed via a formal mathematical framework.
Finally, the work of~\cite{angergaard2022realizing} presents a real-world practical implementation of PA-DFDs that were only conceptual up to that moment, in a Java-based system. The authors demonstrate the enforcement of the privacy constraints in an actual software, and develop two algorithms that generate Java code from PA-DFDs.
%
%This group of works, combines diagrams with formal verification and real-world applications, however most methods are still theoretical or require additional tool support for implementation.  

Beyond PA-DFDs, the work in~\cite{pedroza2021model} uses DFDs to capture data and data flows in their suggested methodology for privacy-aware modelling along with other types of diagrams. Furthermore, ~\cite{veseli2019engineering} uses DFDs through the LINDDUN method to recognize privacy threats in the system model in a real-world case study for an identity management system. Finally, Ferreyra et al., in~\cite{ferreyra2020pdp} propose PDP-ReqLite, a modelling approach introducing the `Requirements Data-Flow Diagram (RDFD)', a DFD-like diagram for requirements-level data flow and processing, and the auxiliary `Personal Information Diagram' (PID) to model personal data properties, enabling automated GDPR privacy requirement generation.

\paragraph{$\bullet$ \textbf{UML Class Diagrams.}} In the different studies, 
Class diagrams are used both in their traditional form to model GDPR roles like the \emph{data subjects}, \emph{processors} or \emph{controllers}
and their relationships, and in extended variants with privacy features. UML-based constraints have been used to automate GDPR compliance verification.
%
%The works in~\cite{torre2020model} and~\cite{torre2021modeling} use 9 class diagrams and 35 OCL constraints to create the generic model of the GDPR, that captures the GDPR's key concepts and their relationships including data subjects, controllers, processors, and processing purposes. 
%
The authors of~\cite{ahmadian2018extending,pedroza2021model} focus on privacy-oriented UML profiles for classifying data and controlling access. More specifically, the work in~\cite{pedroza2021model} defines a privacy-oriented UML profile with constraints about data protection during the design phase, and extends the class diagrams to identify and label %privacy-related entities like the <<DataController>> or 
personal data. %The work also uses the Object Constraint Language (OCL) rules to enforce GDPR constraints during modeling. 
Similarly,~\cite{ahmadian2018extending} uses privacy modelling profiles and extends class diagrams with privacy-related annotations to model and verify the enforcement of privacy-rules. The work in~\cite{ahmadian2019privacy} also extends UML class diagrams with privacy attributes, via \textit{UML Profiles} and \textit{Stereotypes}.

Formal approaches for consent modelling are presented in~\cite{peyrone2022formal2} aiming to provide validation. 
In~\cite{ye2023mbipv}, class diagrams are used 
to describe objects and the relationship between them, establishing the data
model of the proposed methodology for formal validation of privacy properties - in particular, identification of privacy violations.
%In~\cite{peyrone2022formal,peyrone2022formal2} a UML-based consent management framework is developed, with sequence diagrams representing actions related to consent (granting, revoking, and modifying consent). 
%
%There are also works offering real-world applications in certain domains, like %~\cite{veseli2019engineering}. 
%In turn, ~\cite{veseli2019engineering}, uses class diagrams to define data protection requirements for an identity management system. Privacy controls like anonymization and user consent are modeled. 

\paragraph{$\bullet$ \textbf{UML Sequence Diagrams.}} Sequence diagrams are primarily used %in their traditional form 
to demonstrate the data flow in GDPR-compliant system designs. %, for automating GDPR compliance verification.%, and have also been extended with specific GDPR constraints. 
In~\cite{cambronero2024towards} the authors use sequence diagrams to represent cloud service interactions, including consent validation but also data access. 
%Following in specific application domains, ~\cite{veseli2019engineering} uses sequence diagrams to model authentication and consent workflows in identity management systems.
%
%The authors in~\cite{garcia2022contributions} use sequence diagrams to show how data controllers and processors handle user data over time.
%
Additionally,~\cite{vanezi2019gdpr} presents a case study on a web platform, using UML sequence diagrams to model GDPR related functionalities.
On the other hand, the work in~\cite{vanezi2025privacy} extends sequence diagrams presenting the Purpose-Aware Sequence Diagrams (PA-SD) with notation able to describe the ownership, and detailed exchange and use of personal data capturing the processing purposes allowed for each. The work presents a direct mapping between the PA-SD and a formal language created for the same objective.

\paragraph{$\bullet$ \textbf{UML Use Case Diagrams.}} 
Use case diagrams are mainly used both in their traditional form to model privacy-preserving system requirements such as in 
%, like user authentication or consent management as in~\cite{mai2018modeling}. 
the work of~\cite{mougiakou2017based}, which presents the real-world case of an e-learning platform, using use case diagrams for visualising the allocation of GDPR functionalities of the system between actors, and for depicting the data controller and processor with their  GDPR-related responsibilities. 
On the other hand, works extended the use case diagrams, such as in~\cite{ye2023mbipv}, in which the authors employ use case diagrams to model data access control requirements within their proposed methodology to automatically identify privacy violations. Moreover, the authors in~\cite{vanezi2025privacy}, extend use case diagrams with additional notation to describe the personal data input, output, and ownership in a system - providing a high-level overview of purpose-aware requirements. 
%Use case diagrams have also been extended to demonstrate privacy threats in%~\cite{sion2018interaction} and
%~\cite{wuyts2020linddun}. 
%Initially, use case diagrams were used to represent GDPR-related user interactions or system functionalities involving personal data processing, and then `Misuse Case Diagrams' were used to illustrate potential threats to data privacy. 

\paragraph{$\bullet$ \textbf{UML Activity Diagrams.}} In the works studied in this review, activity diagrams are used to visualise workflows related to privacy, to model compliance decision-making processes, and have been extended with GDPR-specific rules. 
%
%Another work in~\cite{veseli2019engineering} uses activity diagrams to model user authentication, identity verification, and privacy settings management, showing the interactions between the actors relating to privacy. 
The work in~\cite{torre2021modeling} uses activity diagrams to model privacy-aware business processes.
The authors in~\cite{ye2023mbipv} developed a methodology and tool to automatically identify privacy violations via formal models, and they use activity diagrams to establish their data flow model. The methodology can identify compliance violations at different stages. Automation in privacy compliance checks is presented in~\cite{ahmadian2018extending}, in which activity diagrams are used to model data sharing in industrial data spaces. 
%
%Lastly, the work in~\cite{torre2021modeling} utilises activity diagrams to model data processing and user interactions and defines decision points where privacy rules are applied.

\paragraph{$\bullet$ \textbf{UML Collaboration Diagrams.}} Collaboration diagrams are used in ~\cite{kammuller2019designing} to show how objects interact to implement the different use cases for the illustrated case study of an IoT healthcare application, incorporating privacy-related functionality. 

% \paragraph{$\bullet$ \textbf{Business Process Modelling Notation (BPMN).}} Pullonen et al.~\cite{pullonen2019privacy} propose the Privacy-Enhanced BPMN (PE-BPMN), an extension of BPMN that introduces privacy-enhancing technology stereotypes and analysis techniques to model and analyse personal data flows and information disclosure in business processes. They do not explicitly address any specific GDPR articles, but PbD in general.

%\paragraph{$\bullet$ \textbf{Goal Models}} Goal models have been employed to 

\subsection{RQ2 – GDPR Notions Addressed}
\label{rq2}

%\noindent\rule{\linewidth}{0.4pt}
\fbox{\begin{minipage}{0.45\textwidth}
\textit{\textbf{RQ2.} Which GDPR privacy notions have been incorporated into software system requirements and design with the use of SE diagrams% UML diagrams or DFDs
?}
\end{minipage}}
\\
\\
A range of GDPR notions are addressed in the studies explored, with the main ones being Privacy-by-Design, Purpose Limitation, Explicit and Informed Consent, the Right to Erasure, Data Minimisation, the Right to Access, Accountability, and Storage Limitation.
Most studies integrate multiple principles, with Purpose Limitation and Privacy-by-Design appearing in nearly all works.
Below, we discuss these notions in turn.

\paragraph{$\bullet$ \textbf{Privacy-by-Design (Article 25 GDPR).}} This principle mandates that privacy should be integrated into systems from the early phases of their design. All 18 works included in this review support Privacy-by-Design (PbD), as GDPR considerations are incorporated during the design phase through the use of diagrams. The studies either model privacy-related functionality and interactions or embed privacy in extended diagrammatic models. In several cases, formal verification is also employed, while in others privacy risks and violations are identified at design time through diagrammatic analysis.

\paragraph{$\bullet$ \textbf{Purpose Limitation (Article 5-1b GDPR).}} This principle requires that personal data only be processed for legitimate purposes, explicitly specified, and agreed upon with the user during data collection. Numerous studies use UML diagrams and DFDs to define, verify, and enforce purpose-specific constraints~\cite{pedroza2021model},~\cite{mougiakou2017based},~\cite{rahman2017petri},~\cite{ahmadian2019privacy}, ~\cite{angergaard2022realizing}, ~\cite{alshareef2021transforming},~\cite{alshareef2021refining}. 
In~\cite{mougiakou2017based} the data controller is assigned, through diagrams, the task of ensuring compliance with GDPR principles relating to the processing of personal data.

The authors in~\cite{alshareef2022precise} propose a methodology in which purpose labels are added to track and verify data usage in system design, building on the work of ~\cite{antignac2018privacy} aiming to ensure that unnecessary data flows are eliminated.
Similarly, the work of~\cite{kammuller2019designing}, uses diagrams to design, monitor, and ensure that data processing is only done for intended purposes. 
Additionally,~\cite{ye2023mbipv} uses diagrams to detect unauthorized data access, %and~\cite{wuyts2020linddun} to identify risks in data usage. 
and~\cite{veseli2019engineering} uses diagrams to model the retrieval of data by users, and recognise relevant threats in terms of purpose of using the data. The authors in~\cite{ahmadian2018extending}, use diagrams to verify if a piece of personal data is processed for the authorized purposes. In the work of~\cite{cambronero2024towards} the methodology uses diagrams to track the requests for data access in cloud environments. The work in~\cite{vanezi2025privacy} focuses on purpose, moving beyond the simple text labels capturing purposes as the allowed use and exchange of personal data as modelled in their suggested Purpose-Aware extended sequence diagrams. Finally, in~\cite{ferreyra2020pdp}, purpose is implicitly addressed, as data flows are tied to processes modelled via Requirements Data-Flow Diagrams and Personal Information Diagrams. Purposes can be inferred, but there is no explicit purpose analysis.

\paragraph{$\bullet$ \textbf{Explicit \& Informed Consent (Articles 6-7 GDPR).}} These articles require that data subjects be clearly informed and provide consent before their personal data are processed. Several works use diagrams to model and verify consent-related interactions~\cite{peyrone2022formal2,torre2021modeling}, while 
%~\cite{wuyts2020linddun} uses diagrams to analyze consent-related threats, and 
~\cite{cambronero2024towards} use diagrams for validating consent prior to any data processing. The work of~\cite{pedroza2021model} visualises information related to consent such as the responsible data controller, the relevant data subject, the process for which the consent is demanded, the purpose, and the personal data involved. The works of~\cite{vanezi2019gdpr} and ~\cite{mougiakou2017based} demonstrate consent modelling, in real-world case studies.

\paragraph{$\bullet$ \textbf{Right to Erasure (Article 17 GDPR).}} 
The Right to Erasure dictates that the data subject can ask and obtain without undue delay the erasure of their personal data from a system. In~\cite{kammuller2019designing}, diagrams are used to demonstrate the deletion process when the right to erasure is exercised by the data subject, while
~\cite{ahmadian2018extending} deals with the notions of retention period through auto-deletion rules. On the more practical side,~\cite{vanezi2019gdpr} and ~\cite{mougiakou2017based} demonstrate how to design such functionality in real-world scenarios.

\paragraph{$\bullet$ \textbf{Data Minimisation (Article 5-1c GDPR).}} The Data Minimisation principle requires that systems collect and process only personal data strictly needed for their purposes. The work in~\cite{torre2021modeling} models system functionality compliant with this principle, while %, or enforce rules for data minimization~\cite{alshareef2021transforming} the work of
~\cite{pedroza2021model} apply design strategies aimed to limiting as much as possible the processing of personally identifiable information.

\paragraph{$\bullet$ \textbf{Right to Access (Articles 15 GDPR).}} Right to Access mandates that data subjects can at any time ask and obtain access to all their personal data stored in a system. %Data portability gives the additional right to the data subject to receive their data in a standard, universal format, able to move them to another similar system.
For example,~\cite{vanezi2019gdpr} demonstrates how access functionality can be designed in a web platform case study using diagrams.

\paragraph{$\bullet$ \textbf{Accountability (Articles 5-2, 24 GDPR).}}
This principle requires that organisations keep auditable logs.
%traceable data processing structures are modeled through diagrams,
%Policies of data processing are modeled in~\cite{ahmadian2019privacy}%, while in~\cite{ahmadian2018extending}, diagrams are used to track data usage logs. 
In~\cite{rahman2017petri}, privacy properties including Accountability are formally verified by Petri net models representing PA-DFDs, while in~\cite{alshareef2021transforming}, PA-DFD elements represent the functionality necessary to enforce accountability.

\paragraph{$\bullet$ \textbf{Storage Limitation (Articles 5-1e GDPR).}} This principle mandates that personal data should be kept for no longer than is necessary for the purposes for which the personal data are processed. The authors in~\cite{ferreyra2020pdp} address Storage Limitation by associating an explicit retention time with personal data elements in their models. 

\subsection{Summary of Cross-Findings} 

Table~\ref{tab:SLR_summary1} provides an overview of the overall results, while Tables~\ref{tab:RQsummary} and~\ref{tab:heat_map} combine the findings from RQ1 and RQ2 by summarising the relationship between diagram types and GDPR notions addressed. Across these results, the Data Flow family of diagrams (DFDs and Privacy-Aware DFDs) dominates, followed by Class diagrams, with other UML diagrams being used less frequently.
%provide summarization by compiling both RQs, with the first one summarizing both the diagram type and GDPR notion addressed by each work, and the latter summarizing how each diagram type corresponds to specific GDPR principles. Finally,
Table~\ref{tab:Works_summary} presents a consolidated view of all studies included in this review.

\begin{table}[ht]
\caption{Overall summary of SLR results.}
\centering
\resizebox{0.47\textwidth}{!}{%
\begin{tabular}{ll}
\hline \textbf{Total Papers Included:} & 18 papers\\
\hline \textbf{Diagram Types Used:} &  
\makecell[l]{DFDs/PA-DFDs, Class Diagrams,\\ Sequence Diagrams, Use Case Diagrams,\\ Activity Diagrams, Collaboration Diagrams}
\\
\hline \textbf{GDPR Notions Addressed:} & 
\makecell[l]{Privacy-by-Design, Purpose Limitation,\\ Explicit \& Informed Consent, Right to Erasure, \\  Data Minimisation, Right to Access,\\ Accountability, Storage Limitation}
\\ \hline
\end{tabular}%
}
\label{tab:SLR_summary1}
\end{table}

As summarized in the tables, the majority of the works (n = 9) employ DFDs, all of which address the notion of Purpose Limitation, with one additionally addressing Consent, two Accountability, and one Storage Limitation.
Class diagrams are used in five studies, mainly to address Purpose Limitation (three works), and additionally Consent (one work), the Right to Erasure (two works), and Data Minimisation (one work). 
Sequence diagrams appear in three studies: one addressing only Purpose Limitation, one addressing Purpose Limitation and Consent, and one addressing Consent, the Right to Erasure, and the Right to Access. 
Use Case diagrams are also employed in three works, all of them addressing Purpose Limitation, with one additionally addressing  Consent and the Right to Erasure. 
Activity diagrams are used in two works, both addressing Purpose Limitation, with one also addressing the Right to Erasure. Finally,  Collaboration diagrams are used in one work, addressing Purpose Limitation and the Right to Erasure.

%%%%%%%
\begin{table*}[ht]
\caption{A compiled summarisation of the results for both RQs, marking the type of diagrams used %(D1: DFDs, D2: Class Diagrams, D3: Sequence Diagrams, D4: Use Case Diagrams, D5: Activity Diagrams, D6: Collaboration Diagrams), 
and the GDPR notions addressed (PbD: Privacy-by-Design, Purp.: Purpose Limitation, Cons.: Consent, Erasure: Right to Erasure, Min.: Data Minimisation, Access: Right to Access, Acc.: Accountability, Stor.: Storage Limitation).}
\centering
\resizebox{\textwidth}{!}{%
\begin{tabular}{|c|c|l|l|l|l|l|l|}
\hline
& \multicolumn{7}{c|}{\textbf{Diagram Type}} \\
\hline 
\rowcolor{Gray}
& & 
 \textbf{DFDs / PA-DFDs}
 & 
\textbf{Class Diagrams} & 
  \textbf{Sequence Diagrams}
 & 
 \textbf{Use Case Diagrams}
  & 
  \textbf{Activity Diagrams}
  & 
 \textbf{Collaboration Diagrams}
  \\
  \hline 
\multirow{10}{*}{\rotatebox{90}{\parbox{2.8cm}{\centering\textbf{GDPR Notion}}}}
&PbD  & \makecell[l]{~\cite{rahman2017petri}, \\
~\cite{alshareef2022precise},\\~\cite{pedroza2021model},\\
~\cite{veseli2019engineering}, \\ ~\cite{antignac2018privacy},\\
~\cite{angergaard2022realizing}, \\
~\cite{alshareef2021refining},\\
~\cite{alshareef2021transforming},\\
~\cite{ferreyra2020pdp}}
& 
 \makecell[l]{~\cite{pedroza2021model}, \\~\cite{ahmadian2018extending}, \\ 
~\cite{peyrone2022formal2}, \\
~\cite{ye2023mbipv},\\
~\cite{ahmadian2019privacy}}
& 
 \makecell[l]{~\cite{cambronero2024towards}, \\ ~\cite{vanezi2019gdpr},\\
 ~\cite{vanezi2025privacy}}
&  \makecell[l]{~\cite{mougiakou2017based}, \\ ~\cite{ye2023mbipv},\\
~\cite{vanezi2025privacy}} &  \makecell[l]{~\cite{ye2023mbipv},\\ ~\cite{ahmadian2018extending}} & ~\cite{kammuller2019designing}\\   \cline{2-8}
&Purp. & 
\makecell[l]{~\cite{rahman2017petri},\\ ~\cite{alshareef2022precise},\\
~\cite{pedroza2021model},\\
~\cite{veseli2019engineering},\\ 
~\cite{antignac2018privacy},\\
~\cite{angergaard2022realizing},\\
~\cite{alshareef2021refining},\\
~\cite{alshareef2021transforming},\\
~\cite{ferreyra2020pdp}}
& \makecell[l]{~\cite{ahmadian2018extending},\\ ~\cite{ye2023mbipv},\\
~\cite{ahmadian2019privacy}}
& \makecell[l]{~\cite{cambronero2024towards}, \\~\cite{vanezi2025privacy}}& \makecell[l]{~\cite{mougiakou2017based},\\
~\cite{ye2023mbipv},\\
~\cite{vanezi2025privacy}
}
&\makecell[l]{~\cite{ye2023mbipv},\\~\cite{ahmadian2018extending}}
&~\cite{kammuller2019designing}\\   \cline{2-8} 
&Cons.  & \makecell[l]{~\cite{pedroza2021model}}  &  \makecell[l]{~\cite{peyrone2022formal2}} & \makecell[l]{~\cite{cambronero2024towards},\\
~\cite{vanezi2019gdpr}}
& ~\cite{mougiakou2017based}& &\\ \cline{2-8}
&Erasure &  & \makecell[l]{~\cite{ahmadian2018extending}, \\~\cite{ahmadian2019privacy}} & ~\cite{vanezi2019gdpr} & ~\cite{mougiakou2017based} & ~\cite{ahmadian2018extending}&~\cite{kammuller2019designing}\\   \cline{2-8}
&Min. &  &~\cite{pedroza2021model} & &  &&\\   \cline{2-8} 
&Access &  & & ~\cite{vanezi2019gdpr} &  &&\\   \cline{2-8} 
&Acc.  & \makecell[l]{~\cite{rahman2017petri},\\ ~\cite{alshareef2021transforming}} &  &  &  &&\\   \cline{2-8} 
&Stor.  & \makecell[l]{~\cite{ferreyra2020pdp}} &  &  &  &&\\ \hline 
\end{tabular}%
}
\label{tab:RQsummary}
\end{table*}

Examining the results from the perspective of GDPR notions addressed, Purpose Limitation emerges as the most frequently modelled GDPR principle (n = 16), with most studies using DFDs. However, all diagram types are used for this principle across different works. Fewer works (n = 5) address Consent using four different diagram types. The Right to Erasure is addressed by five studies using five different diagram types, Data Minimisation by one study using Class diagrams, the Right to Access by one study using Sequence diagrams, Storage Limitation by one study using DFD-like diagrams, and Accountability by two studies using DFDs. 
All 18 studies address Privacy-by-Design, as GDPR compliance is embedded at the design stage.

A closer quantitative examination on Tables~\ref{tab:RQsummary} and~\ref{tab:heat_map} reveals clear patterns in how diagram types are used to address specific GDPR notions. Data Flow Diagrams and Privacy-Aware DFDs are predominantly employed to model Purpose Limitation, reflecting their suitability for representing data-centric concerns and tracking personal data flows. They are also the only diagram family used to represent Storage Limitation.
Sequence diagrams are most frequently used to address Consent, while Class diagrams are the most frequent in addressing the Right to Erasure and the only diagram type observed to model Data Minimisation. 
Sequence diagrams are the only diagrams used to model the Right to Access.

%%%%%%
\begin{table}[h!]
\centering
\caption{Frequency of GDPR notions modelled across diagram types. Darker cells indicate stronger research presence.}
\rowcolors{2}{gray!5}{gray!15}
\resizebox{0.47\textwidth}{!}{%
\begin{tabular}{lcccccccc}
\rowcolor{gray!30}
\textbf{Diagram Type} & PbD & Purp. & Cons. & Erasure & Min. & Access & Acc.& Stor.\\
DFDs / PA-DFDs & \cellcolor{orange!100!white}9 & \cellcolor{orange!100!white}9 & \cellcolor{orange!15!white}1 & \cellcolor{white}0 & \cellcolor{white}0 & \cellcolor{white}0 & \cellcolor{orange!30!white}2&
\cellcolor{orange!15!white}1\\
Class Diagrams & 
\cellcolor{orange!80!white}6 & \cellcolor{orange!45!white}3 & \cellcolor{orange!15!white}1 & \cellcolor{orange!30!white}2 & \cellcolor{orange!15!white}1 & \cellcolor{white}0 & \cellcolor{white}0& 
\cellcolor{white}0\\
Sequence Diagrams & \cellcolor{orange!45!white}3 & \cellcolor{orange!30!white}2 & \cellcolor{orange!30!white}2 & \cellcolor{orange!15!white}1 & \cellcolor{white}0 & \cellcolor{orange!15!white}1 & \cellcolor{white}0& \cellcolor{white}0\\
Use Case Diagrams & \cellcolor{orange!45!white}3 & \cellcolor{orange!45!white}3 & \cellcolor{orange!15!white}1 & \cellcolor{orange!15!white}1 & \cellcolor{white}0 & \cellcolor{white}0 & \cellcolor{white}0& \cellcolor{white}0\\
Activity Diagrams & \cellcolor{orange!30!white}2 & \cellcolor{orange!30!white}2 & \cellcolor{white}0 & \cellcolor{orange!15!white}1 & \cellcolor{white}0 & \cellcolor{white}0 & \cellcolor{white}0& \cellcolor{white}0\\
Collaboration Diagrams & \cellcolor{orange!15!white}1 & \cellcolor{orange!15!white}1 & \cellcolor{white}0 & \cellcolor{orange!15!white}1 & \cellcolor{white}0 & \cellcolor{white}0 & \cellcolor{white}0& \cellcolor{white}0\\
\end{tabular}
}
\smallskip
\resizebox{0.47\textwidth}{!}{%
\begin{tabular}{@{}cccccccccc@{}}
\multicolumn{10}{c}{\small Number of papers} \\
\cellcolor{orange!0!white}\strut\hspace{0.35cm}0\hspace{0.35cm} &
\cellcolor{orange!15!white}\hspace{0.6cm} &
\cellcolor{orange!30!white}\hspace{0.6cm} & 
\cellcolor{orange!45!white}\hspace{0.6cm} &
\cellcolor{orange!60!white}\hspace{0.6cm} &
\cellcolor{orange!70!white}\hspace{0.6cm} & 
\cellcolor{orange!80!white}\hspace{0.6cm} &
\cellcolor{orange!90!white}\hspace{0.6cm} &
\cellcolor{orange!95!white}\strut\hspace{0.6cm}&
\cellcolor{orange!100!white}\strut\hspace{0.25cm}9
\hspace{0.35cm} 
\end{tabular}
}
\label{tab:heat_map}
\end{table}

Overall, the results show that even though privacy requirements are being embedded in modelling notations, most works focus on early-phase design modelling, without providing traceability between privacy requirements and implementation artifacts
or proposing automated tool-supported GDPR compliance checking.
The choice of diagram type appears to be driven by the modelling focus: data-centric diagrams such as DFDs and PA-DFDs are predominantly used to reason about personal data flows such as Purpose Limitation, while UML-based diagrams are more often employed to capture structural aspects, interactions, and user-facing requirements such as Consent and data subject rights.
Furthermore, the reviewed works emphasise Purpose Limitation, while principles such as the Right to Access, Data Minimisation, and Accountability remain under-explored. 
These gaps are discussed further in Section~\ref{discussion_section}.

% %%
% \begin{figure}[th]
% \centering
% \includegraphics[width=0.48\textwidth]{figures/heat-map.png}
% \caption{Frequency of GDPR notions modeled across diagram types. Darker cells indicate stronger research presence.}
% \label{fig:heat-map}
% \end{figure}
% %%

\begin{table*}[ht]
\caption{Overview of the primary studies, indicating diagram types used and GDPR notions addressed.}
\centering
\resizebox{\textwidth}{!}{%
\begin{tabular}{|l|c|c|c|c|c|c|c|c|c|c|c|c|c|c|}
\hline
& \multicolumn{6}{c|}{\textbf{Diagram Type}} & \multicolumn{8}{c|}{\textbf{GDPR Notion}} \\
\hline
 & 
\begin{turn}{90}\makecell[l]{DFDs / \\ PA-DFDs}\end{turn} 
 & 
\begin{turn}{90}\makecell[l]{Class \\Diagrams}\end{turn}  
 & 
\begin{turn}{90}\makecell[l]{Sequence \\Diagrams}\end{turn}
 & \begin{turn}{90}\makecell[l]{Use Case \\Diagrams}\end{turn}
  & \begin{turn}{90}\makecell[l]{Activity \\Diagrams}\end{turn}
  & \begin{turn}{90}\makecell[l]{Collaboration \\Diagrams}\end{turn}
  & \begin{turn}{90}\makecell[l]{Privacy-\\ by-Design}\end{turn} &\begin{turn}{90}\makecell[l]{Purpose \\ Limitation}\end{turn}  & \begin{turn}{90}\makecell[l]{Consent}\end{turn} & \begin{turn}{90}\makecell[l]{Right to \\ Erasure}\end{turn} & \begin{turn}{90}\makecell[l]{Data\\ Minimisation}\end{turn} & \begin{turn}{90}\makecell[l]{Right to\\ Access}\end{turn} &\begin{turn}{90}\makecell[l]{Accountability}\end{turn} &\begin{turn}{90}\makecell[l]{Storage \\Limitation}\end{turn} 
  \\
  \hline
%%%
\cite{ahmadian2018extending}  &  & \checkmark &  &  & \checkmark & & \checkmark & \checkmark & & \checkmark& & &&\\   \hline
  %%%
\cite{ahmadian2019privacy}  &  & \checkmark &  &  &&& \checkmark& \checkmark& &\checkmark & & &&\\   \hline
  %%%
\cite{alshareef2021refining}  & \checkmark  &  &  &  &&& \checkmark & \checkmark & & & & &&\\   \hline
%%%
\cite{alshareef2021transforming}  & \checkmark &  &  &  &&& \checkmark & \checkmark& & & & &\checkmark&\\   \hline
%%%
\cite{alshareef2022precise}  & \checkmark &  &  &  &&& \checkmark & \checkmark & & & & &&\\   \hline
%%%
\cite{angergaard2022realizing}  & \checkmark &  &  &  &&& \checkmark & \checkmark & & & & &&\\   \hline
%%%
\cite{antignac2018privacy}   & \checkmark &  &  &  &&& \checkmark & \checkmark & & & & &&\\   \hline
%%%
\cite{cambronero2024towards}  &  &  & \checkmark &  &&& \checkmark & \checkmark & \checkmark & & & &&\\   \hline
%%%
\cite{kammuller2019designing} & 
& & & &&\checkmark& \checkmark& \checkmark & & \checkmark & & &&\\   \hline 
  %%%
\cite{mougiakou2017based} & 
& & & \checkmark &  & & \checkmark & \checkmark & \checkmark & \checkmark & & &&\\   \hline
%%%
\cite{pedroza2021model} & \checkmark & \checkmark &  &  &&& \checkmark & \checkmark& \checkmark & & \checkmark & &&\\   \hline 
%%%
\cite{peyrone2022formal2} &  & \checkmark & 
& & && \checkmark & & \checkmark & & & &&\\ \hline
%%%
\cite{rahman2017petri}  & \checkmark &  &  &  &&& \checkmark& \checkmark & & & & & \checkmark&\\   \hline 
%%%
\cite{vanezi2019gdpr} &  & & \checkmark&  &&& \checkmark& & \checkmark & \checkmark & & \checkmark &&\\   \hline 
%%%
\cite{veseli2019engineering}  & \checkmark &  &  &  &&& \checkmark & \checkmark & & & & &&\\   \hline 
%%%
\cite{ye2023mbipv}  &  & \checkmark  &  & \checkmark &\checkmark&& \checkmark &\checkmark & & & & &&\\   \hline
%%%
\cite{ferreyra2020pdp} & \checkmark &   &  &  &&& \checkmark & & & & & & &\checkmark\\   \hline
\cite{vanezi2025privacy} &  &   & \checkmark  & \checkmark &&& \checkmark & \checkmark & & & & & &\\   \hline
\end{tabular}%
}
\label{tab:Works_summary}
\end{table*}

\section{Discussion} 
\label{discussion_section}
Building on the findings presented in Section 4, this section discusses the gaps and opportunities revealed. Finally, the section lists the threats to validity of this SLR.

\subsection{Gaps and Opportunities}
\label{gaps}
Our findings reveal several limitations across the reviewed studies.

Studies using DFDs or their privacy-aware extended versions generally restrict modelling to the early system analysis phases, limiting full lifecycle traceability between requirements, design, and implementation, unlike UML models, which can participate in model-driven workflows.
Most approaches focus on isolated design-level representations, without providing explicit links between privacy requirements, design models, and runtime or implementation-level artifacts. The identification of traceability as a gap therefore emerges directly from the absence of such mechanisms across the analysed primary studies.

At the same time, studies lack integration among diagram types, for example, linking class and sequence diagrams, or connecting data flow and use case diagrams, to provide a complete representation of privacy requirements across system design. 

Another important aspect is that most works do not integrate their diagram extensions into actual design environments, and do not provide additional tool support. At the same time, no automated tool-supported compliance checking against GDPR principles is provided.

Finally, there is an evident gap in coverage of the GDPR principles. Modelling efforts focus mainly on Purpose Limitation and Consent, while other principles such as Data Minimisation, the Right to Access, or Accountability are less addressed. 
This focus is understandable: most principles, especially GDPR user rights, are straightforward to represent (e.g., the Right to Erasure can simply be modelled as data deletion functionality). In contrast, notions like Purpose Limitation and Consent are inherently more complex to represent as they depend on each system's specific intended collection, use, and storage of personal data and are tightly coupled with its functional requirements.
Their correct interpretation depends not only on structural elements but also on the purpose, timing, and conditions under which personal data is collected and used.
At the same time, when Purpose Limitation is
addressed, most of the works deal with it via annotations, textual labels, or access control, without dealing with the underlying actions and data flows that implement those purposes.

These gaps highlight opportunities for future research in three key directions:
(1) developing inter-diagram integration and full lifecycle traceability mechanisms;
(2) developing tools to support automated compliance checking, and integrate modelling approaches into mainstream design environments; and
(3) addressing complex GDPR notions, such as Purpose Limitation with more elaborate action-aware modelling approaches, beyond simple textual labelling.

\subsection{Threats to Validity}
We now discuss threats to validity across commonly recognised categories in SLRs:
\begin{enumerate}
    \item Although we used a rigorous search strategy across major literature data sources, it is possible that relevant studies were missed due to limitations in indexing or keyword usage. To mitigate this risk, we complemented the database search with snowballing from the references of included works.
    \item The selection and screening process involved subjective judgment, which may have introduced bias. To reduce this threat, the two reviewing authors independently screened all retrieved studies and discussed any disagreements until joint consensus was reached, with the third author's intervention.
    \item Extracted data might have been interpreted differently by the two reviewers. This risk was mitigated by the first reviewer using a standardized extraction spreadsheet, with cross-checking by the second reviewer.
    \item Our review focused on peer-reviewed academic studies. Grey literature or unpublished industrial experiences may not have been captured.
    %\item The relatively small number of primary studies (n=17) and their concentration on specific diagram types and GDPR notions limit the extent to which the observed trends can be generalized to all privacy-aware system design approaches. Nevertheless, the aim of this review is not statistical generalization but analytical insight into existing research patterns and gaps, which can inform and guide future research directions.
\end{enumerate}

\section{Conclusions}
\label{conclusions_section}
In this work, we presented a Systematic Literature Review on how Software Engineering 
Diagrams are employed to capture and integrate GDPR-based privacy requirements into software system design. Following a rigorous review protocol, we collected and analysed 18 primary studies published between 2017 and 2025, addressing two research questions on: (i) the types of diagrams employed, and (ii) the GDPR notions addressed. We also recognised key gaps and opportunities based on the findings. The findings revealed the need for inter-diagram integration, full lifecycle traceability mechanisms, tool-supported automated compliance checking, and environment integration.

\balance

\bibliographystyle{apalike}
{\small
\bibliography{example}}

\end{document}